\documentclass[paper,12pt]{elsarticle}

\usepackage{amssymb}
\begin{document}

\begin{frontmatter}



\title{Universal Fluctuations of AEX index}

 \author{Rui Gon\c calves}
 \address{LIAAD-INESC Porto LA and Faculty of Engineering, University of Porto. R. Dr. Roberto Frias s/n, 4200-465 Porto, Portugal.}
  \ead{rjasg@fe.up.pt}

\author{Helena Ferreira \fnref{label4}}
\address{LIAAD-INESC Porto LA, Portugal}
\ead{helenaisafer@gmail.com}
\fntext[label4]{Corresponding Author}
 
\author{Alberto Pinto}
\address{LIAAD-INESC Porto LA and  Department of Mathematics, Faculty of Sciences, University of Porto. Rua do Campo Alegre, 687, 4169-007, Portugal.}
\ead{aapinto1@gmail.com}

%
%uthor{Rui Gon\c calves $^{\rm a}$, Helena Ferreira $^{\rm b}$ and Alberto Pinto $^{\rm c}$, \corref{cor1}}
%%\vspace{6pt} $^{\rm a}${\em{LIAAD-INESC Porto LA and Faculty of Engineering, University of Porto, R. Dr. Roberto Frias s/n, 4200-465 Porto, Portugal}}\\
%%$^{\rm b}$ {\em LIAAD-INESC Porto LA, Portugal}\\
%%$^{\rm c}$ {\em LIAAD-INESC Porto LA and  Department of Mathematics, Faculty of Sciences, University of Porto. Rua do Campo Alegre, 687, 4169-007, Portugal}
%
%\cortext[cor1]{aapinto1@gmail.com}
\begin{abstract}
We compute the analytic expression of the probability distributions $F_{AEX,+}$ and $F_{AEX,-}$ of the normalized positive and negative AEX (Netherlands) index daily returns $r(t)$. Furthermore, we  define the $\alpha$ re-scaled AEX daily index positive returns  $r(t)^\alpha$ and negative returns $(-r(t))^\alpha$ that we call, after normalization, the $\alpha$ positive fluctuations and $\alpha$ negative  fluctuations. We use the Kolmogorov-Smirnov statistical test, as a method, to find the values of  $ \alpha$ that optimize the data collapse of the histogram of the $ \alpha$  fluctuations  with the   Bramwell-Holdsworth-Pinton (BHP) probability density function. The optimal parameters that we found are $\alpha^{+}= 0.46$  and $\alpha^{-}= 0.43$. Since the BHP probability density function appears in several other dissimilar phenomena, our result reveals an universal feature of the stock exchange markets.
\end{abstract}


\end{frontmatter}

% \linenumbers

\section{Introduction}
The modeling of the time series of stock  prices is a main issue in
economics and finance and it is of a vital importance in the
management of large portfolios of stocks \cite{Gabaixetal03, LilloMan01, ManStan95}.  Here, we
analyze the AEX Index. The AEX index, derived from Amsterdam Exchange index, is a stock market index composed of Dutch companies that trade on Euronext Amsterdam, formerly known as the Amsterdam Stock Exchange. Started in 1983, the index is composed of a maximum of 25 of the most actively traded securities on the exchange. It is one of the main national indices of the pan-European stock exchange group Euronext alongside Brussels' BEL20, Paris's CAC 40 and Lisbon's PSI-20. The time series to investigate in our analysis is the \emph{ AEX index} from October of 1992 to October of 2009.
 Let $Y(t)$ be the AEX index adjusted close value at day $t$. We define the \emph{AEX index daily return} on day $t$ by
$$
r(t)=\frac{Y(t)-Y(t-1)}{Y(t-1)}.
$$
We define the $\alpha$ \emph{re-scaled AEX daily index positive returns} $r(t)^\alpha$, for $r(t)>0$, that we call, after normalization, the $\alpha$ \emph{positive fluctuations}. We define the $\alpha$ \emph{re-scaled AEX daily index negative returns} $(-r(t))^\alpha$, for $r(t)<0$, that we call, after normalization, the $\alpha$  \emph{negative fluctuations}.
We analyze, separately, the $\alpha$ positive and $\alpha$ negative daily fluctuations that can have different statistical and economic natures due, for instance, to the leverage effects (see, for example, \cite{Andersen, Barnhartetal09, Pinto, Pinto1}).
Our aim is to find the values of $\alpha$ that optimize the data collapse of the histogram of the $\alpha$ positive and  $\alpha$ negative
fluctuations to the universal, non-parametric, Bramwell-Holdsofworth-Pinton (BHP) probability density function.
To do it, we apply the Kolmogorov-Smirnov statistic test to the null hypothesis claiming
that the probability distribution of the $\alpha$
fluctuations is equal to the (BHP) distribution.
We observe that the $P$ values of the Kolmogorov-Smirnov test
vary continuously with $\alpha$. The highest $P$ values  $P^{+}=0.59...$ and $P^{-}=0.31...$ of the Kolmogorov-Smirnov test are attained for the values $\alpha^{+}= 0.46...$  and $\alpha^{-}= 0.43...$, respectively, for the positive and negative fluctuations. Hence, the null hypothesis is not rejected for values of $\alpha$ in  small neighborhoods of $\alpha^{+}= 0.46...$  and $\alpha^{-}= 0.43...$.
Then, we show the data collapse of the histograms of the $\alpha^{+}$ positive fluctuations and $\alpha^{-}$ negative fluctuations to the  BHP pdf. Using this data collapse, we do a change of variable that allow us to compute the analytic expressions of the probability density 
 functions $f_{AEX,+}$ and $f_{AEX,-}$
 of the normalized positive and negative  AEX index daily returns
\begin{eqnarray*}
f_{AEX,+}(x) &=& 4.39x^{-0.54}f_{BHP}(20.28x^{0.46}-2.12) \\
f_{AEX,-}(x) &=& 3.52x^{-0.57}f_{BHP}(17.53x^{0.43}-2.14)
\end{eqnarray*}
in terms of the BHP pdf $f_{BHP}$. 
We exhibit the data collapse of the histogram of the positive and negative returns to 
our proposed theoretical pdf´s $f_{AEX,+}$ and $f_{AEX,-}$.
Similar results are
observed for some other stock indexes, prices of stocks, exchange rates and
commodity prices (see \cite{Gonc, Gond}).
Since the BHP probability density function appears in several other dissimilar phenomena (see, for instance, \cite{bramwellfennelleuphys2002,
DahlstedtJensen2001, DahlstedtJensen2005, Gona, Gonb, Gonf, Gong, Pinto}), our result reveals an universal feature of the stock exchange markets. 

\section{Positive AEX index daily returns}
\begin{figure}[htbp!] 
\begin{center}
\includegraphics[width=8cm]{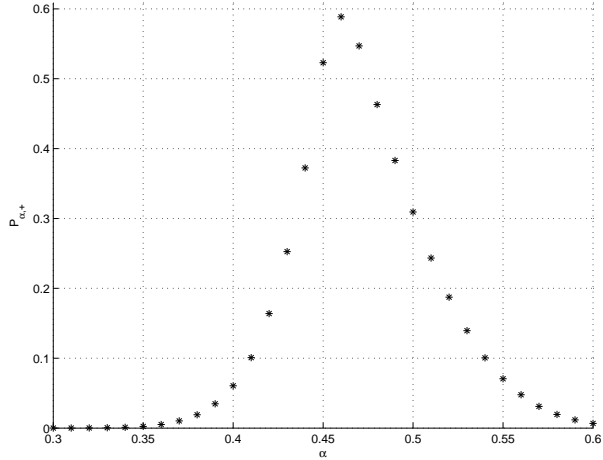}
\caption{\footnotesize{The Kolmogorov-Smirnov $P$ value $P_{\alpha,+}$ for values of $\alpha$ in the range $[0.30, 0.60]$. 
}} \label{fig1}
\end{center}
\end{figure}

Let $T^+$ be the set of all days $t$ with positive returns, i.e.
 $$
 T^+=\{t:r(t)>0\} .
 $$
 Let $n^+=2285$ be the cardinal of the set $T^+$. The \emph{$\alpha$ re-scaled AEX daily index positive returns} are the returns $r(t)^\alpha$ with $t\in T^+$. Since the total number of observed days is $n=4318$, we obtain that  $n^+/n=0.53$.
 The \emph{mean} $\mu^+_{\alpha}=0.105...$ of the $\alpha$ re-scaled AEX daily index positive returns  is given by
\begin{equation}
\mu^+_{\alpha}=\frac{1}{n^{+}}\sum_{t\in T^+}r(t)^\alpha
 \label{eq2}
\end{equation}
The \emph{standard deviation}  $\sigma^+_{\alpha}=0.049...$ of the $\alpha$ re-scaled AEX daily index positive returns  is given by
\begin{equation}
\sigma^+_{\alpha}=\sqrt{\frac{1}{n^{+}}\sum_{t\in T^+} {r(t)^{2\alpha}} - (\mu^+_{\alpha})^2}
 \label{eq3}
\end{equation}
\noindent
We define the $\alpha$ \emph{positive fluctuations} by
\begin{equation}
r^+_{\alpha}(t) = \frac{r(t)^\alpha - \mu^+_{\alpha}}{\sigma^+_{\alpha}}
 \label{eq6}
\end{equation}
\noindent
for every $t\in T^+$. Hence, the $\alpha$ \emph{positive fluctuations} are the normalized $\alpha$ re-scaled $AEX$ daily index positive returns.
Let $L^+_{\alpha}=-1.97...$ be the \emph{smallest} $\alpha$ positive fluctuation, i.e.
$$
L^+_{\alpha}=\min_{t\in T^+}\{r^+_{\alpha}(t)\}.
$$
Let $R^+_{\alpha}=5.08...$ be the \emph{largest} $\alpha$ positive fluctuation, i.e.
$$
R^+_{\alpha}=\max_{t\in T^+}\{r^+_{\alpha}(t)\}.
$$
We denote by $F_{\alpha,+}$ the \emph{probability distribution of the $\alpha$ positive fluctuations}.
Let the \emph{truncated BHP probability distribution} $F_{BHP,\alpha,+}$ be given by
$$
F_{BHP, \alpha,+}(x)=\frac{F_{BHP}(x)}{F_{BHP}(R^+_{\alpha})-F_{BHP}(L^+_{\alpha})}
$$
where $F_{BHP}$ is the BHP probability distribution.
We apply the Kolmogorov-Smirnov statistic test to the null hypothesis claiming that the probability distributions $F_{\alpha,+}$ and $F_{BHP,\alpha,+}$ are equal. The Kolmogorov-Smirnov $P$ \emph{value} $P_{\alpha,+}$  is  plotted in Figure \ref{fig1}.
Hence, we observe that $\alpha^+=0.46...$ is the point where the $P$ value $P_{\alpha,+} =0.59...$  attains its maximum.
 
\noindent

\begin{figure}[htbp!]
\begin{center}
\includegraphics[width=8cm]{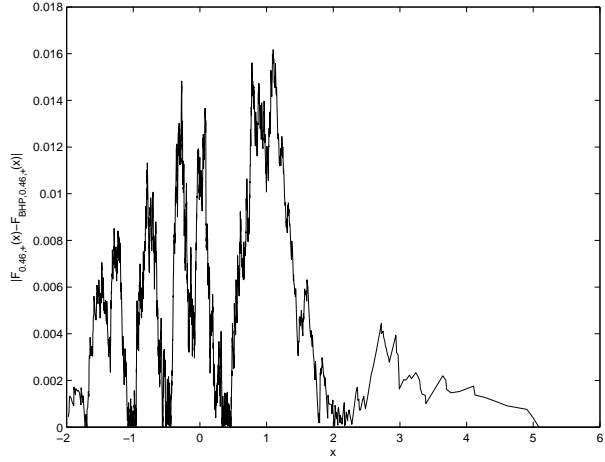}
\caption{\footnotesize{The map $D_{0.46,+}(x)=|F_{0.46,+}(x)-F_{BHP,0.46,+}(x)|$.}}
 \label{fig2}
\end{center}
\end{figure}

\noindent 
It is well-known that the Kolmogorov-Smirnov $P$ value $P_{\alpha,+}$  decreases with the distance  
$\left\|F_{\alpha,+}-F_{BHP,\alpha,+}\right\|$
between $F_{\alpha,+}$ and $F_{BHP,\alpha,+}$.
In Figure \ref{fig2}, we plot $D_{\alpha^+,+}(x)=\left|F_{\alpha^+,+}(x)-F_{BHP,\alpha^+,+}(x)\right|$ and we observe that 
$D_{\alpha^+,+}(x)$ attains its highest values for the $\alpha^+$ positive fluctuations  above or close to the mean of the probability distribution.\\

\begin{figure}[htbp!]
\begin{center}
\includegraphics[width=8cm]{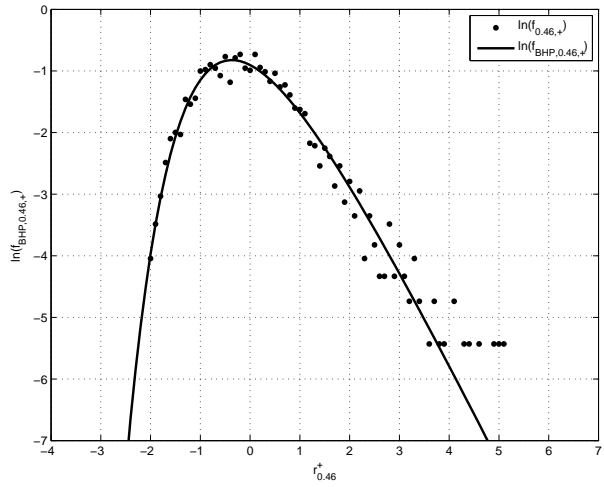}
\caption{\footnotesize{The histogram  of the $\alpha^+$ positive fluctuations with the truncated BHP  pdf $f_{BHP,0.46,+}$ on top, in the semi-log scale.}}
 \label{fig3}
\end{center}
\end{figure}

\begin{figure}[htbp!]
\begin{center}
\includegraphics[width=8cm]{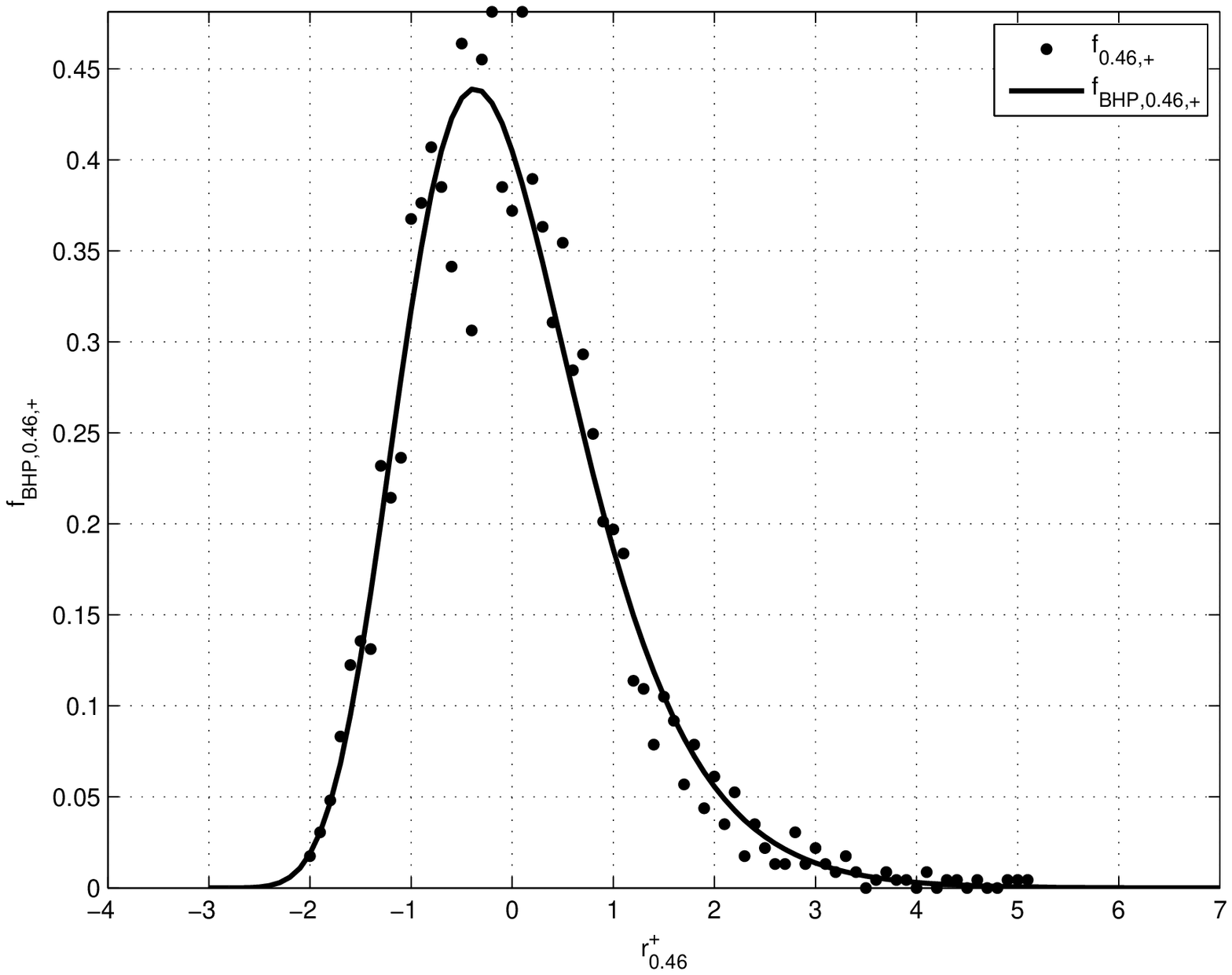}
\caption{\footnotesize{The histogram  of the $\alpha^+$ positive fluctuations with the truncated BHP pdf $f_{BHP,0.46,+}$ on top.}}
 \label{fig4}
\end{center}
\end{figure}

\noindent In Figures \ref{fig3} and \ref{fig4}, we show the data collapse of the histogram $f_{\alpha^+,+}$ of the $\alpha^+$ positive fluctuations to the  truncated BHP pdf  $f_{BHP,\alpha^+,+}$. \\

\noindent Assume that the probability distribution of the $\alpha^+$ positive fluctuations $r^+_{\alpha^+}(t)$ is given by $F_{BHP,\alpha^+,+}$ (see \cite{Gonb}).
The pdf $f_{AEX,+}$ of the AEX daily index positive returns $r(t)$ is given by
$$
f_{AEX,+}(x)= \frac{\alpha^+ x^{\alpha^+-1}f_{BHP}\left(\left(x^{\alpha^+}-\mu^+_{\alpha^+}\right)/\sigma^+_{\alpha^+}\right)}{\sigma^+_{\alpha^+}\left(F_{BHP}\left(R^+_{\alpha^+}\right)-F_{BHP}\left(L^+_{\alpha^+}\right)\right)}.
$$

\begin{figure}[htbp!]
\begin{center}
\includegraphics[width=8cm]{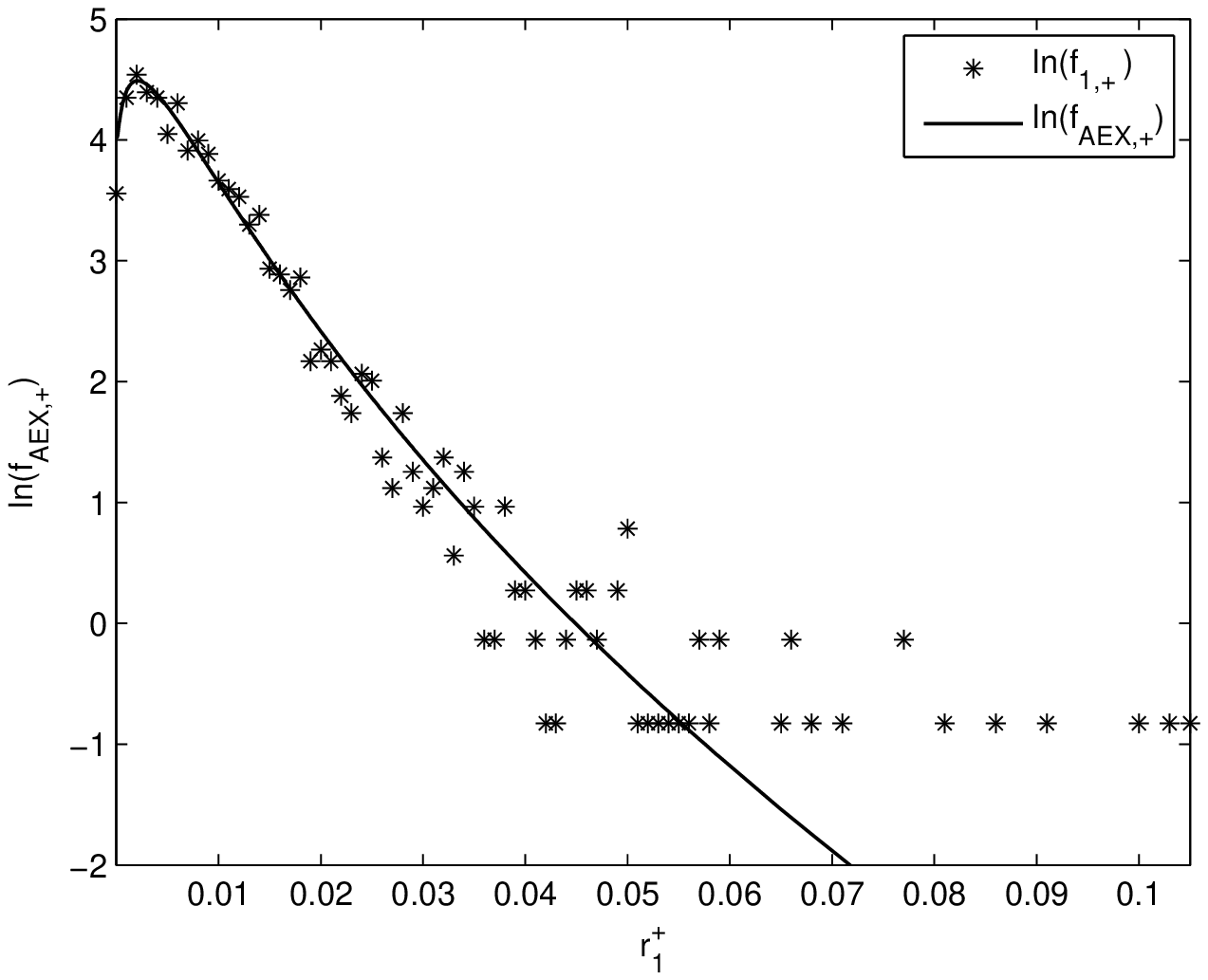}
\caption{\footnotesize{The histogram of the fluctuations
 of the positive returns with the pdf $f_{AEX,+}$ on top, in the semi-log scale.}}
 \label{fig5}
\end{center}
\end{figure}

\begin{figure}[htbp!]
\begin{center}
\includegraphics[width=8cm]{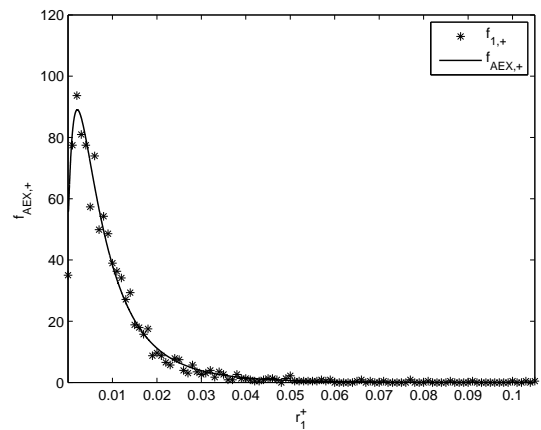}
\caption{\footnotesize{The histogram of the fluctuations
 of the positive returns with the pdf $f_{AEX,+}$ on top.}}
 \label{fig6}
\end{center}
\end{figure}

\noindent
Hence, taking $\alpha^+=0.46...$, we get
$$
f_{AEX,+}(x)=4.39...x^{-0.54...}f_{BHP}(20.28...x^{0.46...}-2.12...).
$$
In Figures \ref{fig5} and \ref{fig6}, we show the data collapse of the histogram $f_{1,+}$ of the positive returns to 
our proposed theoretical pdf $f_{AEX,+}$. 

\noindent

\section{Negative AEX index daily returns}

\begin{figure}[htbp!]
\begin{center}
\includegraphics[width=8cm]{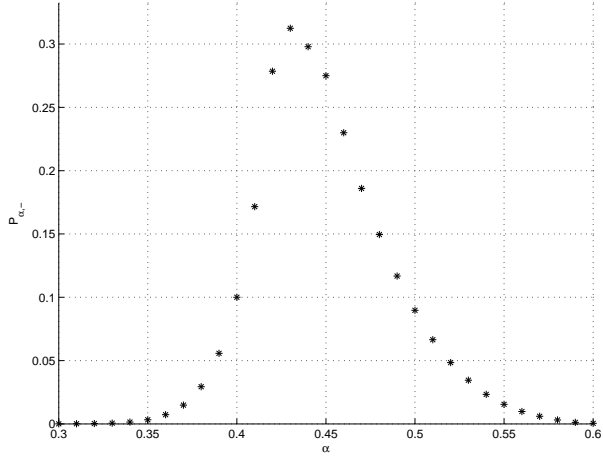}
\caption{\footnotesize{The Kolmogorov-Smirnov $P$ value $P_{\alpha,-}$ for values of $\alpha$ in the range $[0.30, 0.60]$.
}} \label{fig7}
\end{center}
\end{figure}

\begin{figure}[htbp!]
\begin{center}
\includegraphics[width=8cm]{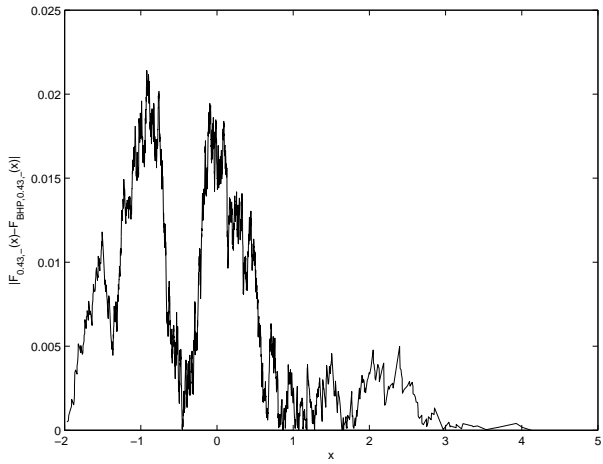}
\caption{\footnotesize{The map $D_{0.43,-}(x)=|F_{0.43,-}(x)-F_{BHP,0.43,-}(x)|$.}} \label{fig8}
\end{center}
\end{figure}

Let $T^-$ be the set of all days $t$ with negative returns, i.e.
 $$
 T^-=\{t:r(t)<0\} .
 $$
 Let $n^-=2021$ be the cardinal of the set $T^-$.  Since   the total number of observed days is $n=4318$, we obtain that  $n^-/n=0.47$.
 The \emph{$\alpha$ re-scaled AEX daily index negative returns} are the returns $(-r(t))^\alpha$ with $t\in T^-$. We note that  $-r(t)$ is positive.
 The \emph{mean} $\mu^-_{\alpha}=0.122...$ of the $\alpha$ re-scaled AEX daily index negative returns  is given by
\begin{equation}
\mu^-_{\alpha}=\frac{1}{n^{-}}\sum_{t\in T^-}(-r(t))^\alpha
 \label{eq2}
\end{equation}
The \emph{standard deviation}  $\sigma^-_{\alpha}=0.057...$ of the $\alpha$ re-scaled AEX daily index negative returns  is given by
\begin{equation}
\sigma^-_{\alpha}=\sqrt{\frac{1}{n^{-}}\sum_{t\in T^-} {(-r(t))^{2\alpha}} - (\mu^-_{\alpha})^2}
 \label{eq3}
\end{equation}
\noindent
We define the $\alpha$ \emph{negative fluctuations} by
\begin{equation}
r^-_{\alpha}(t) = \frac{(-r(t))^\alpha - \mu^-_{\alpha}}{\sigma^-_{\alpha}}
 \label{eq6}
\end{equation}
\noindent
for every $t\in T^-$. Hence, the $\alpha$ \emph{negative fluctuations} are the normalized $\alpha$ re-scaled $AEX$ daily index negative returns.
Let $L^-_{\alpha}=-1.97...$ be the \emph{smallest} $\alpha$ negative fluctuation, i.e.
$$
L^-_{\alpha}=\min_{t\in T^-}\{r^-_{\alpha}(t)\}.
$$
Let $R^-_{\alpha}=4.13...$ be the \emph{largest} $\alpha$ negative fluctuation, i.e.
$$
R^-_{\alpha}=\max_{t\in T^-}\{r^-_{\alpha}(t)\}.
$$
We denote by $F_{\alpha,-}$ the \emph{probability distribution of the $\alpha$ negative fluctuations}.
Let the \emph{truncated BHP probability distribution} $F_{BHP,\alpha,-}$ be given by
$$
F_{BHP, \alpha,-}(x)=\frac{F_{BHP}(x)}{F_{BHP}(R^-_{\alpha})-F_{BHP}(L^-_{\alpha})}
$$
where $F_{BHP}$ is the BHP probability distribution.
We apply the Kolmogorov-Smirnov statistic test to the null hypothesis claiming that the probability distributions $F_{\alpha,-}$ and $F_{BHP,\alpha,-}$ are equal. 
The Kolmogorov-Smirnov $P$ \emph{value} $P_{\alpha,-}$  is  plotted in Figure \ref{fig7}. Hence, we observe that $\alpha^-=0.43...$ is the point where the $P$ value $P_{\alpha^-} =0.31...$  attains its maximum. 
\noindent
\noindent
The Kolmogorov-Smirnov $P$ value $P_{\alpha,-}$  decreases with the distance  
$\left\|F_{\alpha,-}-F_{BHP,\alpha,-}\right\|$
between $F_{\alpha,-}$ and $F_{BHP,\alpha,-}$.
In Figure \ref{fig8}, we plot $D_{\alpha^-,-}(x)=\left|F_{\alpha^-,-}(x)-F_{BHP,\alpha^-,-}(x)\right|$ and we observe that 
$D_{\alpha^-,-}(x)$ attains its highest values for the $\alpha^-$ negative fluctuations below the mean of the probability distribution.\\

\begin{figure}[htbp!]
\begin{center}
\includegraphics[width=8cm]{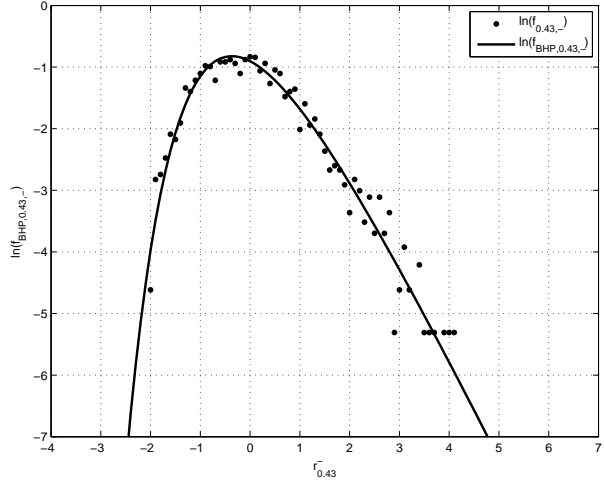}
\caption{\footnotesize{The histogram of the $\alpha^-$ negative fluctuations
 with the truncated BHP pdf $f_{BHP,0.43,-}$ on top, in the semi-log scale.}}
 \label{fig9}
\end{center}
\end{figure}

\begin{figure}[htbp!]
\begin{center}
\includegraphics[width=8cm]{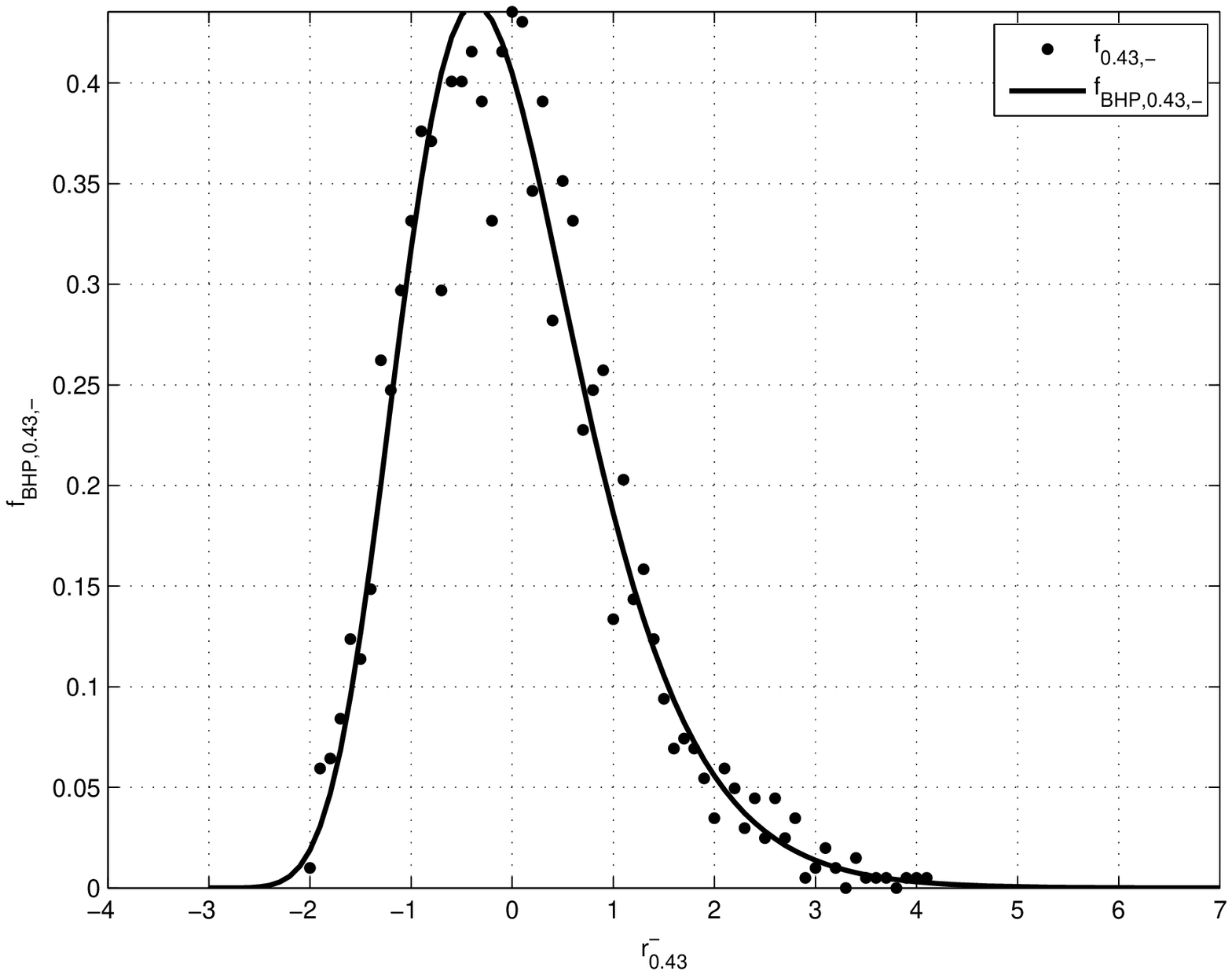}
\caption{\footnotesize{The histogram of the $\alpha^-$ negative fluctuations
 with the truncated BHP pdf $f_{BHP,0.43,-}$ on top.}}
 \label{fig10}
\end{center}
\end{figure}

\noindent In Figures \ref{fig9} and \ref{fig10}, we show the data collapse of the histogram $f_{\alpha^-,-}$  of the $\alpha^-$ negative fluctuations to the truncated BHP pdf $f_{BHP,\alpha^-,-}$. \\

\noindent Assume that the probability distribution of the $\alpha^-$ negative fluctuations $r^-_{\alpha^-}(t)$ is given by $F_{BHP,\alpha^-,-}$, (see \cite{Gonb}).
The pdf $f_{AEX,-}$ of the AEX daily index (symmetric) negative returns $-r(t)$, with $T \in T^-$,  is given by
$$
f_{AEX,-}(x)=  \frac{\alpha^- x^{\alpha^-1}f_{BHP}\left(\left(x^{\alpha^-}-\mu^-_{\alpha^-}\right)/\sigma^-_{\alpha^-}\right)}{\sigma^-_{\alpha^-}\left(F_{BHP}\left(R^-_{\alpha^-}\right)-F_{BHP}\left(L^-_{\alpha^-}\right)\right)}.
$$

\noindent
Hence, taking $\alpha^-=0.43...$, we get
$$
f_{AEX,-}(x)=3.52...x^{-0.57...}f_{BHP}(17.53...x^{0.43...}-2.14...)
$$

\begin{figure}[htbp!]
\begin{center}
\includegraphics[width=8cm]{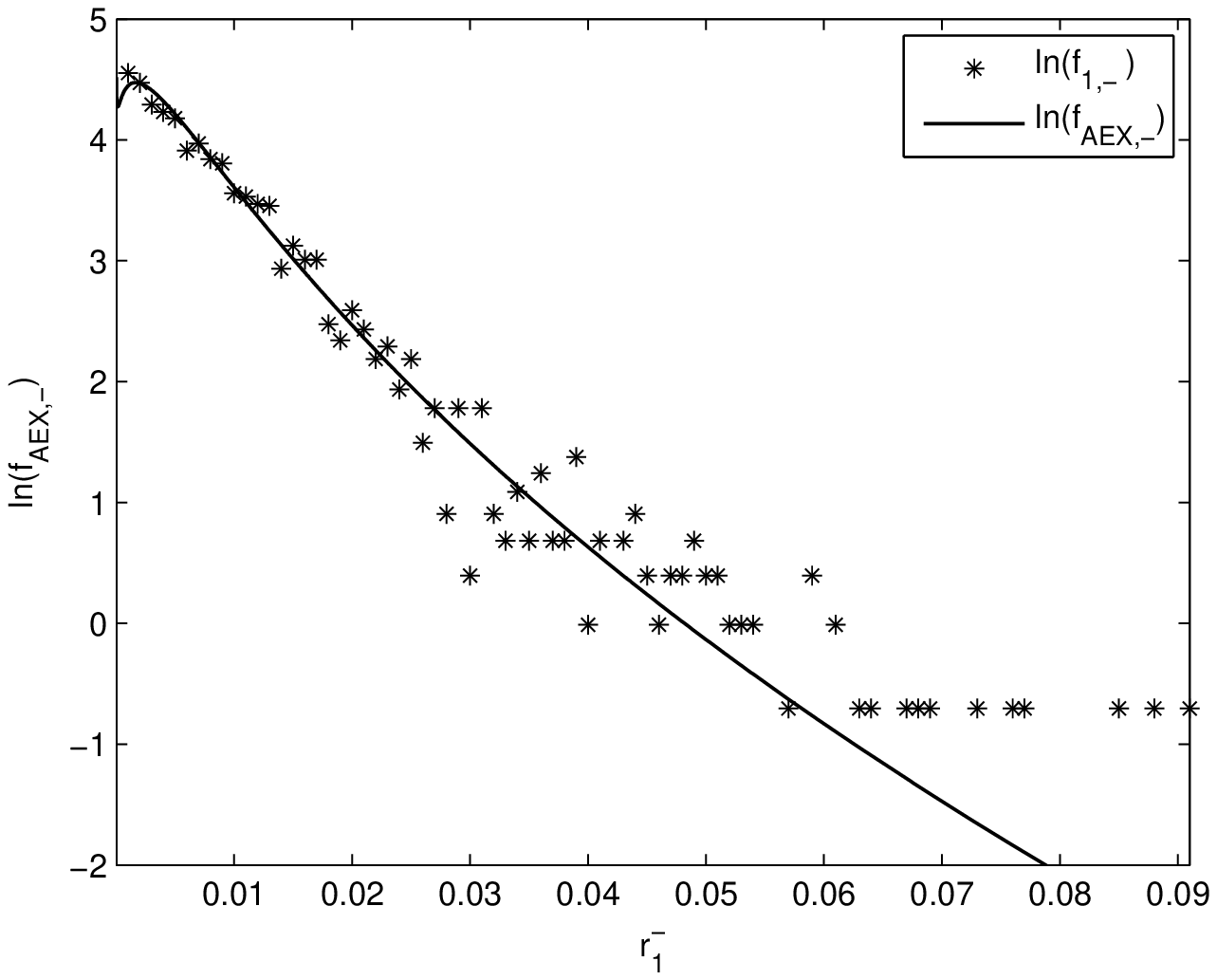}
\caption{\footnotesize{The histogram of the negative returns with the pdf $f_{AEX,-}$ on top, in the semi-log scale.}}
 \label{fig11}
\end{center}
\end{figure}

\begin{figure}[htbp!]
\begin{center}
\includegraphics[width=8cm]{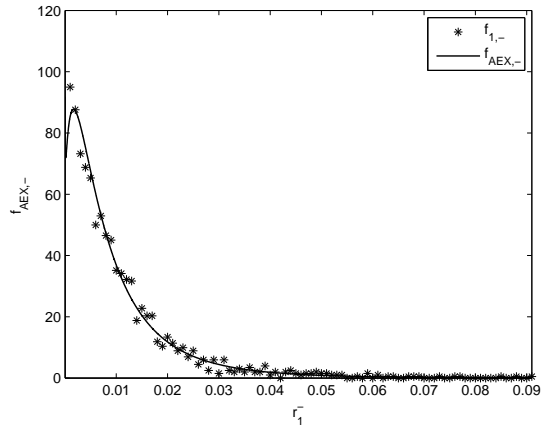}
\caption{\footnotesize{The histogram of the negative returns with the pdf $f_{AEX,-}$ on top, in the semi-log scale.}}
 \label{fig12}
\end{center}
\end{figure}

\noindent In Figures \ref{fig11} and \ref{fig12}, we show the data collapse of the histogram $f_{1,-}$ of the negative returns to 
our proposed theoretical pdf $f_{AEX,-}$.

\section{Conclusions}
 We used the Kolmogorov-Smirnov statistical test to compare the histogram of the $\alpha$ positive fluctuations and $\alpha$ negative
fluctuations with the universal, non-parametric, Bramwell-Holdsworth-Pinton (BHP) probability distribution.
We found that the parameters $\alpha^{+}= 0.46...$  and $\alpha^{-}= 0.43...$ for the positive and negative fluctuations, respectively, optimize the $P$ value of the Kolmogorov-Smirnov test. We obtained that the respective $P$ values of the
Kolmogorov-Smirnov statistical test are $P^{+}=0.59...$ and $P^{-}=0.31...$. Hence, the null hypothesis was not rejected.
The fact that $\alpha^+$ is different from $\alpha^-$ can be do to leverage effects.
We presented the data collapse of the corresponding fluctuations histograms to the BHP pdf.
Furthermore, we computed the analytic expression of the probability distributions $F_{AEX,+}$ and $F_{AEX,-}$ of the normalized AEX index daily positive and negative returns in terms of the BHP pdf. We showed the data collapse of the histogram of the positive and negative returns to 
our proposed theoretical pdfs $f_{AEX,+}$ and $f_{AEX,-}$. The results obtained in daily returns also apply to other periodicities, such as weekly and monthly returns as well as intraday values.

In \cite{Gonc, science},  it is found the 
data collapses of the histograms of some other stock indexes, prices of stocks, exchange rates, 
commodity prices and energy sources \cite{s6} to the BHP pdf.

Bramwell, Holdsworth and Pinton \cite{BHP1998} found the
probability distribution of the fluctuations of the total magnetization,
in the strong coupling (low temperature) regime, for a
two-dimensional spin model (2dXY) using the spin wave
approximation. From a statistical physics point of view,
one can think that the stock prices form a non-equilibrium system
\cite{Chowdhury, Gopikrishnanetal98, LilloMan01, Plerouetal99}.
Hence, the results presented here lead to a construction of a new qualitative and quantitative
econophysics model for the stock market based in the two-dimensional spin
model (2dXY) at criticality (see \cite{Gond}).

\section*{Acknowledgments}
We thank Henrik Jensen, Peter Holdsworth  and Nico Stollenwerk for
showing us the relevance of the Bramwell-Holdsworth-Pinton
distribution.
This work was presented in PODE09, EURO XXIII, Encontro Ci\^encia 2009 and ICDEA2009.
We thank LIAAD-INESC Porto LA, Calouste Gulbenkian Foundation, PRODYN-ESF, POCTI and POSI by FCT and Minist\'erio da Ci\^encia e da Tecnologia, and the FCT Pluriannual Funding Program of the LIAAD-INESC Porto LA. Part of this research was developed during a visit by the authors to the IHES, CUNY, IMPA, MSRI, SUNY, Isaac Newton Institute and University of Warwick. We thank them for their hospitality.

\appendix
\section {Definition of the Bramwell-Holdsworth-Pinton probability distribution}
The universal nonparametric BHP pdf was discovered by Bramwell,
Holdsworth and Pinton \cite{BHP1998}. The \emph{BHP probability density function (pdf)} is
given by
\begin{eqnarray}
             &\!&f_{BHP}(\mu)=\int_{-\infty}^{\infty}\frac{dx}{2\pi}
\sqrt{\frac{1}{2N^2}\sum_{k=1}^{N-1}\frac{1}{\lambda_k^2}}
e^{ix\mu\sqrt{\frac{1}{2N^2}\sum_{k=1}^{N-1}\frac{1}{\lambda_k^2}}}\nonumber\\\!
&\!&
.e^{-\sum_{k=1}^{N-1}\left[\frac{ix}{2N}\frac{1}{\lambda_k}-\frac{i}{2}
\mbox{arctan}\left(\frac{x}{N\lambda_k}\right)\right]}.e^{-\sum_{k=1}^{N-1}\left[\frac{1}{4}\mbox{ln}{\left(1+\frac{x^2}{N^2\lambda_k^2}\right)}\right]}\nonumber\\
      \label{eq1}
\end{eqnarray}
\noindent where the $\{\lambda_k\}_{k=1}^L$ are the eigenvalues, as
determined in \cite{Bramwelletal2001}, of the adjacency matrix. It
follows, from the formula of the BHP pdf, that the asymptotic values
for large deviations, below and above the mean, are exponential and
double exponential, respectively (in this article, we use the
approximation of the BHP pdf obtained by taking $L=10$ and $N=L^2$
in equation (\ref{eq1})). As we can see, the BHP distribution does
not have any parameter (except the mean that is normalize to 0 and
the standard deviation that is normalized to 1) and it is universal,
in the sense that appears in several physical phenomena. For
instance, the universal nonparametric BHP distribution is a good
model to explain the fluctuations of order parameters in theoretical
examples such as, models of self-organized criticality, equilibrium
critical behavior, percolation phenomena (see \cite{BHP1998}), the
Sneppen model (see \cite{BHP1998} and \cite{DahlstedtJensen2001}),
and auto-ignition fire models (see \cite{SinharayBordaJensen2001}).
The universal nonparametric BHP distribution is, also, an
explanatory model for fluctuations of several phenomenon such as,
width power in steady state systems (see \cite{BHP1998}),
fluctuations in river heights and flow (see \cite{Bramwelletal2001, DahlstedtJensen2005,Gona, Gonb, Gonf}), for the plasma density
fluctuations and electrostatic turbulent fluxes measured at the
scrape-off layer of the Alcator C-mod Tokamaks (see
\cite{VanMilligen05}) and for Wolf's sunspot numbers
 fluctuations (see \cite{Gong}).

\bibliographystyle{elsarticle-num}

\begin {thebibliography}{12}

\bibitem{Andersen}
T.G. Andersen, T. Bollerslev, P. Frederiksen and M. Nielse,
Continuos-Time Models, Realized Volatilities and Testable Distributional Implications for Daily Stock Returns
{\it Preprint} (2004).

\bibitem{Barnhartetal09}
S. W. Barnhart and A. Giannetti,
Negative earnings, positive earnings and stock return predictability: An empirical examination of market timing
{\it Journal of Empirical Finance} {\bf 16} 70-86 (2009).


\bibitem{BHP1998}
        S.T. Bramwell, , P.C.W. Holdsworth and J.F. Pinton,
       {\it Nature}, {\bf 396} 552-554 (1998).

\bibitem{bramwellfennelleuphys2002}
        S.T. Bramwell, T. Fennell, P.C.W. Holdsworth and B. Portelli,
        Universal Fluctuations of the Danube Water
        Level: a Link with Turbulence, Criticality and Company Growth,
        {\it Europhysics Letters  }{\bf 57} 310 (2002).

\bibitem{Bramwelletal2001}
        S.T. Bramwell, J.Y. Fortin, P.C.W. Holdsworth, S.
        Peysson, J.F. Pinton, B. Portelli and M. Sellitto,
        Magnetic Fluctuations in the classical XY model:
        the origin of an exponential tail in a complex system,
        {\it Phys. Rev  }{\bf E 63} 041106 (2001).


\bibitem{Chowdhury}
 D. Chowdhury and D. Stauffer,
A generalized spin model of financial markets
{\it Eur. Phys. J.} {\bf B8} 477-482 (1999).

\bibitem{DahlstedtJensen2001}
        K. Dahlstedt and H.J. Jensen,
        Universal fluctuations and extreme-value statistics,
        {\it J. Phys. A: Math. Gen. }{\bf 34} 11193-11200 (2001).

\bibitem{DahlstedtJensen2005}
         K. Dahlstedt and H.J. Jensen, 
        Fluctuation spectrum and size scaling of river flow and level,
        {\it Physica }{\bf A 348} 596-610 (2005).

\bibitem{science} Dynamics, Games and Science. 
Eds: M. Peixoto, A. A. Pinto and D. A.  Rand.
Proceedings in Mathematics series, Springer-Verlag (2010).

\bibitem{Gabaixetal03}
        X. Gabaix, G. Parameswaran, V. Plerou and H. E. Stanley,
        A theory of power-law distributions in financial markets
        {\it Nature} {\bf 423} 267-270 (2003).

\bibitem{Gonb}
      R. Gon\c calves, H. Ferreira and A. A. Pinto, 
      Universality in the Stock Exchange Market
			{\it Journal of Difference Equations and Applications} (accepted in 2009).
			
\bibitem{s6}
R. Gon\c calves, H. Ferreira and A. A. Pinto,
{\em Universality in energy sources},
 IAEE (International Association for Energy economics) International Conference (accepted in 2010).

\bibitem{Gonc}
        R. Gon\c calves, H. Ferreira and A. A. Pinto,
        Universal fluctuations of the Dow Jones
        (submitted).


\bibitem{Gond}
        R. Gon\c calves, H. Ferreira and A. A. Pinto,
        A qualitative and quantitative Econophysics stock market model
        (submitted).
 

\bibitem{Gona}
        R. Gon\c calves, H. Ferreira, A. A. Pinto and N. Stollenwerk,
        Universality in nonlinear prediction of complex systems.
        Special issue in honor of Saber Elaydi. {\it Journal of Difference Equations and Applications} {\bf 15}, Issue 11 \& 12, 1067-1076 (2009).
        
\bibitem{Gonf}
       R. Gon\c calves and A. A. Pinto,
       Negro and Danube are mirror rivers.
      Special issue Dynamics \& Applications in honor of Mauricio Peixoto and David Rand. {\it Journal of Difference Equations and Applications} (2009).


\bibitem{Gong}
        R. Gon\c calves, A. A. Pinto and N. Stollenwerk,
        Cycles and universality in sunspot numbers fluctuations
        {\it The Astrophysical Journal} {\bf 691} 1583-1586 (2009).


\bibitem{Gopikrishnanetal98}
        P. Gopikrishnan, M. Meyer, L. Amaral and H. E. Stanley,
        Inverse cubic law for the distribution of stock price
        variations.
        {\it The European Physical Journal B} {\bf 3} 139-140 (1998).


\bibitem{LilloMan01}
F. Lillo and R. Mantegna,
 Ensemble Properties of securities traded in the Nasdaq market
 {\it Physica A} {\bf 299} 161-167 (2001).



\bibitem{ManStan95}
   R. Mantegna and H. E. Stanley,
   Scaling behaviour in the dynamics of a economic index
 {\it Nature} {\bf 376} 46-49 (2001).


\bibitem{Pinto}
        A. A. Pinto,
        Game theory and Duopoly Models
        {\it Interdisciplinary Applied Mathematics}, Springer-Verlag (2010).
        
\bibitem{Pinto1}        
A. A. Pinto, D. A. Rand and F. Ferreira, 
Fine Structures of Hyperbolic Diffeomorphisms. 
{\it Springer-Verlag Monograph} (2009).

\bibitem{Plerouetal99}
      V. Plerou, L. Amaral, P. Gopikrishnan, M. Meyer and H. E. Stanley,
       Universal and Nonuniversal Properties of Cross Correlations
       in Financial Time Series
       {\it Physical Review Letters} {\bf 83} 7 1471-1474 (1999).

\bibitem{SinharayBordaJensen2001}
        P. Sinha-Ray, L. Borda de \'Agua and H.J. Jensen,
        Threshold dynamics, multifractality and universal fluctuations
        in the SOC forest fire: facets of an auto-ignition model
        {\it Physica }{\bf D 157}, 186--196 (2001).


\bibitem{VanMilligen05}
        B. Ph. Van Milligen, R. S\'anchez, B. A. Carreras, V. E. Lynch,
        B. LaBombard, M. A. Pedrosa, C. Hidalgo, B. Gon\c calves and  R. Balb\' in,
        {\it Physics of plasmas} {\bf 12} 05207 (2005).





\end{thebibliography}

%\listoffigures
%%
%% End of file `elsarticle-template-num.tex'.
\end{document}